\documentclass[aps,pre,twocolumn,superscriptaddress]{revtex4-1}
\usepackage[english]{babel}
\usepackage[utf8x]{inputenc}
\usepackage{amsmath,amsfonts,amssymb,amsbsy,eucal,dsfont}
\usepackage{natbib}
\usepackage{color}
\usepackage{graphicx}
\usepackage{rotating}
\usepackage{setspace}

\begin{document}

\title{Liquid pair correlations in four spatial dimensions:\\Theory versus simulation}

\author{M. Heinen}
\affiliation{Division of Chemistry and Chemical Engineering,
California Institute of Technology,
Pasadena, California 91125, USA.}

\author{J.~Horbach}
\affiliation{Institut f\"{u}r Theoretische Physik II,
Weiche Materie,
Heinrich-Heine-Universit\"{a}t D\"{u}sseldorf,
40225 D\"{u}sseldorf,
Germany}

\author{H.~L\"owen}
\affiliation{Institut f\"{u}r Theoretische Physik II,
Weiche Materie,
Heinrich-Heine-Universit\"{a}t D\"{u}sseldorf,
40225 D\"{u}sseldorf,
Germany}

\begin{abstract}
Using liquid integral equation theory, we calculate the pair correlations of particles that interact via a smooth repulsive pair potential
in $d=4$ spatial dimensions. We discuss the performance of different closures for the Ornstein-Zernike equation,
by comparing the results to computer simulation data. Our results are of relevance to understand crystal and glass formation in high-dimensional systems. 
\end{abstract}

\maketitle

\section{Introduction}\label{sec:intro}

Particle-resolved structure in a classical homogeneous bulk liquid is typically
measured in terms of pair correlation functions. In real space,
the pair correlations provide the conditional probability density to find a particle at a
distance $r$ from another particle. The associated Fourier transform
correlates density waves of wavenumber $k$ \cite{Hansen_McDonald1986}.
While the latter is a typical outcome of a scattering experiment \cite{Westermeier2012},
the former can be obtained from the real-space coordinates of the individual particles
such as colloids \cite{Royall2006,Assoud2009} or dusty plasmas \cite{Book_Ivlev_Royall}.
Computer simulations of classical many-body systems with a prescribed
particle pair-interaction potential are a suitable and well-established route to calculate pair
correlations \cite{AllenTildesley_Book}. There are, however, situations that require
a \mbox{(semi-)analytical} statistical mechanical approach as an alternative to computer simulations.
In such cases, there is a choice of various liquid integral equations that are based on the Ornstein-Zernike
equation, and that have often been proven to predict pair correlations accurately and efficiently
\cite{Hansen_McDonald1986,Nagele1996,Caccamo1996,Parola1995,Brader2008,CarbajalTinoco2008,Heinen2011,Westermeier2012,Heinen2014}.
The large body of complementary experiments, simulations and analytical theory have resulted
in a good understanding of pair correlations in the fluid or liquid state by now.

Most of the studies so far have focused on the physically most relevant situation
of three spatial dimensions ($d=3$), but liquids can also exist in lower
spatial dimensions when they are confined \cite{Alba-Simionesco2006,Lowen2009},
\textit{e.g.}, to two-dimensional interfaces \cite{Zahn1999} or between plates \cite{Schmidt1997}
($d=2$) or inside narrow cylindrical tubes ($d=1$) \cite{Oguz2011}.
Though they do not possess an immediate physical realization, higher spatial dimensions ($d>3$) have been another focus
of recent research. The motivation to consider dimensions higher than three derives from the
ambition to understand the salient necessary ingredients for freezing and the glass transition.
In the existing literature on higher dimensional particulate systems, hard hyperspheres
have mostly been studied \cite{Baus1987,Gonzalez-Melchor2001,Skoge2006,Finken2002,vanMeel2009}
while there are less studies for particles with soft pair potentials like, \textit{e.g.},
the Lennard-Jones potential \cite{Eaves2009, Bruning2009}.

The computational effort for particle-resolved computer simulations rises quickly as a function of $d$,
which effectively limits computer simulations to dimensions $d \lesssim 12$ \cite{Charbonneau2011}.
Analytical \cite{Robles2004, Rohrmann2007, Adda-Bedia2008} or efficient numerical \cite{Heinen2014}
methods for the calculation of pair correlations in dimensions $d>3$ are therefore worth aspiring for.
In the present study, we examine the accuracy of the hypernetted chain (HNC) \cite{Morita1958},
Percus-Yevick (PY) \cite{Percus1958}, and Rogers-Young (RY) \cite{Rogers1984} integral equations,
which are compared with numerically accurate computer simulations of particles with soft interactions
of the Weeks-Chandler-Andersen (WCA) \cite{Weeks1971} type, in $d=4$ spatial dimensions. We find
that (just like in $d=3$ spatial dimensions) the RY scheme predicts pair-correlations in excellent
agreement with the computer simulation results, while the PY and HNC scheme show severe over- and
under-estimation of the undulations in the static structure factor, respectively.

The remaining parts of this article are organized as follows:
In section~\ref{sec:WCA} we define the WCA fluid under study.
We continue in section~\ref{sec:sim} to outline the computer simulations,
and in section~\ref{sec:LIE} the liquid integral equations that
we use to compute particle pair correlations.
Results for the static structure factor are presented in section~\ref{sec:S_of_q}, which is followed by our
concluding remarks given in section~\ref{sec:conclusions}.

\section{Weeks-Chandler-Andersen pair potential}\label{sec:WCA}

We study homogeneous fluids of spherically symmetric, monodisperse particles
that interact via a smoothed Weeks-Chandler-Andersen (WCA) potential \cite{Weeks1971}, \textit{i.e.}, a Lennard-Jones 
potential of depth $\varepsilon$, which has been truncated at the minimum position $r = r_c = 2^{1/6} \sigma$, and shifted upwards by $\varepsilon$. 
Thus, this non-negative (repulsive) pair potential is defined by
\begin{equation}\label{eq:WCApot}
u(r) = \left\lbrace
   \begin{array}{lr}
   0\,& \hspace{-5em} \text{for}~r > r_c = 2^{1/6}\sigma,\\~\\
   f(r) \left[ 4\varepsilon \left( {\left( \dfrac{\sigma}{r} \right)}^{12} - {\left( \dfrac{\sigma}{r} \right)}^{6} \right) + \varepsilon \right]& ~~\text{otherwise},\\
   \end{array}
 \right. \, \\
\end{equation}
where $f(r) = {(r-r_c)}^4 / [ {(\sigma/200)}^4 + {(r-r_c)}^4 ]$ is a smoothing function that decays rapidly
from $f(r) \approx 1$ for $r < r_c - \sigma/200$ to $f(r) = 0$ for $r = r_c$. The function $f(r)$
provides continuity of forces at $r_c$ and thus, in a molecular dynamics simulation, a better
numerical stability is achieved when using this smoothing function. 

The thermodynamic equilibrium state of the WCA fluid studied here is fully described by two dimensionless parameters:
The normalized thermal energy $k_B T / \epsilon$ (with Boltzmann constant $k_B$) and  $n \sigma^d$, which is
the number of particles in a  $d$-dimensional volume $\sigma^d$. Here, $n = N / L^d$ is the number density
for $N$ particles in a hypercubic box of edge length $L$ (taken in the thermodynamic limit $N\to\infty$ and $L\to\infty$, where $n$ is held fixed).
Note that in the limiting case of vanishing temperature ($T \to 0$ or $\epsilon \to \infty$), the smoothed WCA potential in
Eq.~\eqref{eq:WCApot} reduces to the pair-potential of hard spheres with diameter $r_c$.

\section{Computer simulations}\label{sec:sim}

We performed molecular dynamics (MD) simulations of a four-dimensional,
monodisperse system of 20000 particles that interact via the WCA
potential, as given by Eq.~(\ref{eq:WCApot}). Newton's equations of
motion were integrated with the velocity form of the Verlet algorithm
using a time step of $\delta t = 0.00072$ in units of $\tau = \sqrt{m
\sigma^2/\varepsilon}$ (with mass $m = 1.0$). The particles were put into
a simulation box with linear dimension $L=10.511205\,\sigma$, applying
periodic boundary conditions in all four spatial directions. Simulations
were done at the temperatures $T=1.66, 1.7, 1.8, 1.85, 1.9, 2.0, 2.5,
4.0, 7.0$ (in units of $\varepsilon/k_B$). At each temperature, the
system was fully equilibrated, requiring equilibration runs between $10^5$
time steps at $T=7.0\,\varepsilon/k_B$ and $4 \times 10^7$ time steps
at $T=1.66\,\varepsilon/k_B$. The equilibration runs were followed by
production runs of double length, from which the structure factor $S(q)$
was computed. During equilibration, temperature was fixed by periodically
coupling the system to a stochastic heat bath. The production runs were
done in the microcanonical ensemble. Note that none of the runs showed
any sign of crystallization.


\section{Liquid integral equations}\label{sec:LIE}

The Ornstein-Zernike equation for homogeneous and isotropic, $d$-dimensional fluids reads
\begin{equation}\label{eq:OZ}
h(r) = c(r) + n \int d^d r'~c(r') h(r-r') 
\end{equation}
in terms of the $d$-dimensional particle number density
and the total and direct correlation functions $h(r)$ and $c(r)$, respectively \cite{Hansen_McDonald1986}.
To obtain a closed integral equation for a given kind of pair-potential $u(r)$,
the Ornstein-Zernike equation must be supplemented by a closure relation. With the exception
of very small number densities, exact closure relations are unknown in general and one has to resort to
approximate closures. Here we study three different approximate closures. The first two are the PY closure \cite{Percus1958}
\begin{equation}\label{eq:PY}
c(r) = \left[ \gamma(r) + 1 \right] \times \left[ e^{-\beta u(r)} - 1 \right]
\end{equation}
and the HNC closure \cite{Morita1958}
\begin{equation}\label{eq:HNC}
c(r) = - \gamma(r) - 1 + e^{\gamma(r) - \beta u(r)},
\end{equation}
both written in terms of the indirect correlation function $\gamma(r) = h(r) - c(r)$ and the  
inverse thermal energy $\beta = 1/(k_B T)$. 
Both the PY closure and the HNC closure are thermodynamically inconsistent, in the sense that the predicted 
normalized inverse isothermal osmotic compressibility computed in the fluctuation route, 
\begin{equation}\label{equ:invcomp_fluct}
\dfrac{1}{\chi_c} = {\beta \left( \frac{\partial P_c}{\partial n} \right)}_{T} = 1 - n \int\limits_0^\infty c(r) dr,
\end{equation}
does not match the corresponding expression
\begin{equation}\label{equ:invcomp_vir}
\dfrac{1}{\chi_v} = {\beta \left( \frac{\partial P_v}{\partial n} \right)}_{T},
\end{equation}
from the virial route.
Here, $P_v$ is the virial pressure which, for monodisperse particles with WCA pair-potentials as studied here,
can be calculated according to
\begin{equation}\label{equ:vir_press}
\dfrac{\beta P_v}{n} = \left\lbrace
   \begin{array}{ll}
   1 + \dfrac{\beta~n~\omega_d}{2 d}~r_c^d~g(r_c^+)\,& ~~\text{for}~T = 0,\\~\\
   1 - \dfrac{\beta~n~\omega_d}{2 d} \displaystyle{\int\limits_0^\infty dr~r^d~g(r)~\dfrac{d u(r)}{d r}}.\,& ~~\text{for}~T > 0.\\
   \end{array}
 \right. \, \\
\end{equation}
In Eq.~\eqref{equ:vir_press}, $g(r) = h(r) + 1$ is the radial distribution function, $\omega_d = 2 \pi^{d/2} / \Gamma(d/2)$
is the $d$-dimensional unit hypersphere surface in terms of the Gamma function ($\omega_4 = 2\pi^2$), and $g(r_c^+) = \lim_{r\searrow r_c} g(r)$
is the contact value of the hard-sphere radial distribution function in the special case of $T=0$.

Thermodynamic inconsistency with respect to the isothermal compressibility can be
avoided by using the RY closure \cite{Rogers1984} 
\begin{equation}\label{eq:RY}
c(r) = - \gamma(r) - 1 + e^{-\beta u(r)} \left[ 1+ \dfrac{e^{\gamma(r)f(r)} - 1}{f(r)}\right],
\end{equation}
where $f(r) = 1 - \exp\left\lbrace \alpha r \right\rbrace$ is a mixing function that depends on the non-negative inverse length $\alpha$.
The RY closure interpolates between the PY closure (which is recovered in both limits $r \to 0$ and $\alpha \to 0$), and the HNC
closure (recovered for $r \to \infty$ or $\alpha \to \infty$). The parameter $\alpha$ is selected such that equal values
are obtained for the isothermal osmotic compressibility calculated in the fluctuation route and the virial route.
The standard RY scheme, as used in the present work, is thermodynamically self-consistent with respect to the isothermal osmotic
compressibility only. At the expense of an increased numerical effort, the RY scheme can be further improved by requiring
consistency in additional, independent thermodynamic quantities \cite{CarbajalTinoco2008}.  
Note also that the RY scheme usually does not have a solution for non-positive-definite pair potentials. However, for such potentials
different thermodynamically partially self-consistent closure relations have been devised \cite{Zerah1986, Bomont2004}, similar in spirit to the RY scheme. 

\begin{figure}
\begin{center}
\includegraphics[width=.75\columnwidth,angle=-90]{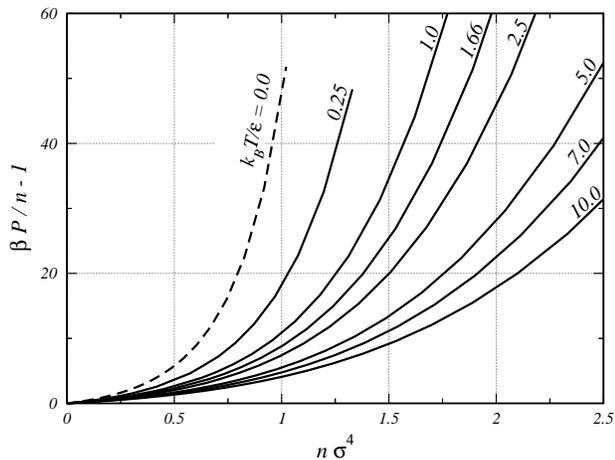}
\caption{\label{fig:EOS}
Equations of state for four-dimensional WCA fluids, for eight different reduced temperatures $k_B T / \epsilon = 0.0, 0.25, 1.0, 1.66, 2.5, 5.0, 7.0,$ and $10.0$, as indicated.
The Rogers-Young normalized excess pressure is plotted as a function of the number of particles in a volume $\sigma^4$.
The dashed curve for $T = 0.0$ is the result for a four-dimensional fluid of hard spheres with diameter $r_c = 2^{1/6} \sigma$. 
}
\end{center}
\end{figure}

The equation of state of four-dimensional WCA fluids at various temperatures is investigated in
Fig.~\ref{fig:EOS}, where we plot the excess part of the normalized pressure, $\beta P / n - 1$, as calculated in the RY scheme.

The (partial) thermodynamic self consistency of the RY scheme usually results in a significantly improved accuracy of the pair-correlation
functions including $g(r)$ and the static structure factor $S(q) = 1 + n \mathcal{F}[h(r)](q)$, with $\mathcal{F}$ denoting the $d$-dimensional
Fourier transform of an isotropic function. However, the good accuracy of the RY scheme is essentially an empirical finding that should be
tested by comparison to simulation results, for each pair-potential and each number of spatial dimensions.

In the important, generic case of hard hyperspheres, the (in this case rather accurate) PY
integral equation \cite{Percus1958} can in principle be solved analytically for arbitrary odd dimension \cite{Robles2004, Rohrmann2007},
and semi-analytically for arbitrary even dimension \cite{Adda-Bedia2008}. However, these \mbox{(semi-)analytical} solution methods are
quite cumbersome, with an analytical effort that rises quickly with increasing number of dimensions $d$.
Moreover, the PY scheme usually over-estimates the undulations in the pair-correlation functions (in particular in the static structure
factor), when it is applied to soft repulsive pair-potentials.
In case of soft particle interactions, like studied in the present work, analytical progress can be made, \textit{e.g.}, by resorting to the 
mean spherical approximation (MSA) \cite{Hansen_McDonald1986} which represents a closure for the Ornstein-Zernike equation that results in a linear integral equation.
Remedy for the unsatisfactory accuracy of the MSA has been proposed in several studies \cite{HansenHayter1982, Snook1992, Heinen2011},
in form of semi-analytical rescaling arguments.

A versatile and computationally efficient alternative to the \mbox{(semi-)analytical} solution of arbitrary-dimensional liquid integral equations
is the numerical solution by means of a spectral solver. Within this numerical method, employed in the present study, it is easy to implement
a variety of different closures for the Ornstein-Zernike equation, suitable for a variety of particle pair-interaction potentials. In the present
work, we employ a numerical method that we have comprehensively outlined in Ref.~\cite{Heinen2014}. This method, based on techniques that were originally
published in Refs.~\cite{Ng1974, Talman1978, Rossky1980, Hamilton2000, Hamilton_website}, is applicable in all positive spatial dimensions $d$
and is numerically very efficient and robust. Our implementation of the numerical solution algorithm allows to compute solutions
for $1 \leq d \lesssim 30$, the upper boundary for $d$ depending on the kind of pair-potentials, the closure relation, and the particle number density.

As a further motivation for studying liquid integral equations in higher dimensions, we note here that
mode-coupling theory (MCT) has been employed to study the glass transition of hard hyperspheres in very high dimensions \cite{Schmid2010}.
In Ref.~\cite{Schmid2010}, the structural input to the MCT equations was generated by approximating $c(r)$ by the Mayer function
$\exp\left\lbrace-\beta u(r)\right\rbrace - 1$. As outlined in Ref.~\cite{Schmid2010}, the
latter approximation is exact in the limit of infinite spatial dimension ($d \to \infty$), and remains to be a good approximation
of the actual particle pair correlations for dimensions $d \gtrsim 100$. For dimensions $d$ in the range $1 \leq d \lesssim 100$,
the approximation $c(r) \approx \exp\left\lbrace-\beta u(r)\right\rbrace - 1$ is insufficient unless the particle number density is very low.
This leads to unphysical artifacts in the
predicted glass transition lines for $d \lesssim 100$ \cite{Schmid2010}. As an alternative, one can use simulation results for the static pair correlations
as input to MCT \cite{Charbonneau2010}. However, the computational effort of computer simulations rises quickly as a function $d$, which limits this approach
to dimensions $d \lesssim 12$ \cite{Charbonneau2011}.
Hence there is a gap for $10 \lesssim d \lesssim 100$, where neither simulation results nor infinite-dimensional limiting expressions
can be used. This gap can be essentially filled in by numerical liquid integral equation solutions as reported in the present article and in Ref.~\cite{Heinen2014}.

\section{Static structure factors}\label{sec:S_of_q}

\begin{figure*}
\begin{center}
\includegraphics[width=.55\textwidth,angle=-90]{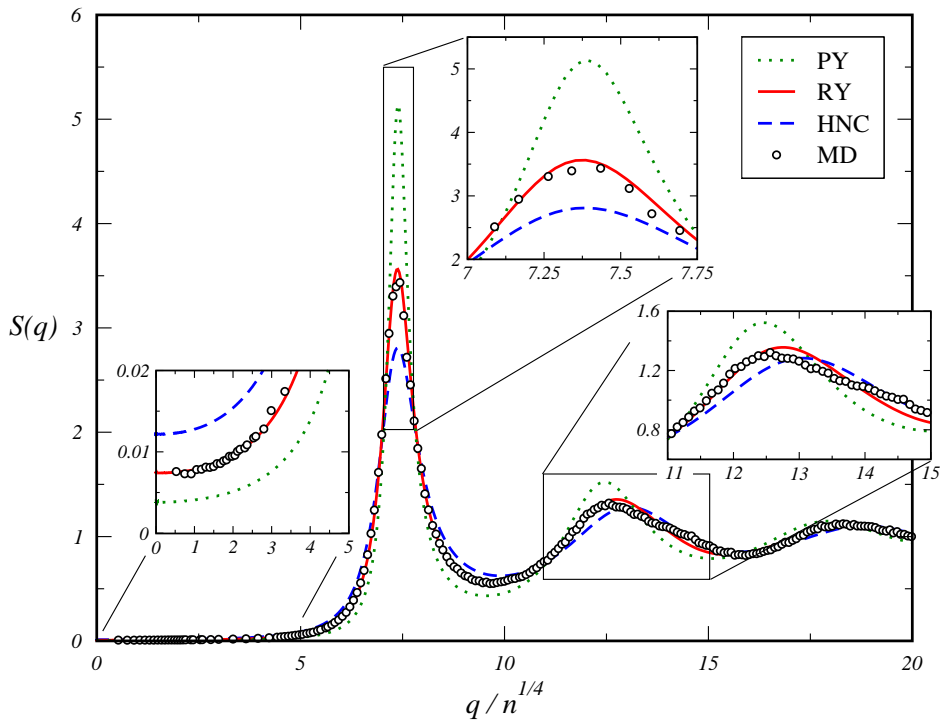}
\caption{\label{fig:Sq_T1_66}
Static structure factor for a four dimensional fluid of particles interacting via WCA pair potentials [Eq.~\eqref{eq:WCApot}],
for particle number density $n = 1.6384 \sigma^{-4}$ and for $k_B T = 1.66\varepsilon$.
Black circles filled in white: Molecular dynamics simulation results.
Dotted green curve: Percus-Yevick scheme.
Solid red curve: Rogers-Young scheme.
Blue dashed curve: Hypernetted chain scheme.
The three insets magnify the region of very low wave numbers $q$,
the region around the principal peak, and around the second peak.
}
\end{center}
\end{figure*}

In Fig.~\ref{fig:Sq_T1_66} we plot the static structure factor, $S(q)$, for a four-dimensional WCA fluid at a rather high number density
$n = 1.6384 \sigma^{-4}$, and a rather low temperature, $T = 1.66 \varepsilon / k_B$. Under these conditions the fluid exhibits very pronounced
pair correlations. Shown are the results from our Molecular Dynamics simulation (black circles filled in white), and from the HNC (blue dashed curve),
RY (red solid curve), and PY (green dotted curve) integral equations. Three insets magnify the regions of very low wave numbers, $q \gtrsim 0$,
the region around the structure factor's principal peak, and the region around the second peak. Note that the HNC scheme predicts a structure factor with considerably
underestimated undulations, and that the PY scheme is severely over-estimating these undulations, while the RY scheme is in very good (if not excellent) agreement
with the simulation result. Each of these observations is in-line with the usual observations that have been made for three-dimensional fluids of purely repulsive particles. 

\begin{figure*}
\begin{center}
\vspace{-4em}
\includegraphics[width=.4\textwidth,angle=-90]{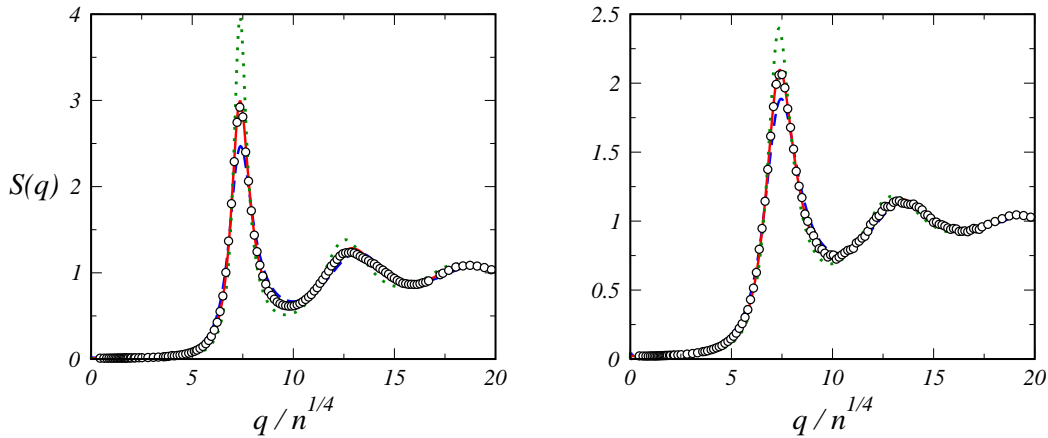}
\caption{\label{fig:Sq_T2_5_and_7_0}
Static structure factors for two four-dimensional fluids of particles interacting via WCA pair potentials [Eq.~\eqref{eq:WCApot}],
for dimension $d=4$, particle number density $n = 1.6384 \sigma^{-4}$, and for $k_B T = 2.5\varepsilon$ (left panel) and $k_B T = 7.0\varepsilon$ (right panel).
Black circles filled in white: Molecular dynamics simulation results.
Dotted green curves: Percus-Yevick scheme.
Solid red curves: Rogers-Young scheme.
Blue dashed curves: Hypernetted chain scheme.
}
\end{center}
\end{figure*}

The only obvious difference between the MD structure factor and the RY structure factor (and the PY and HNC results alike) is a failure of the liquid
integral equation schemes to predict the shape of the second peak in $S(q)$: The right flank of the second peak in the simulation result exhibits
a nearly linear decay of $S(q)$ for values of $q / n^{1/4}$ between $12.5$ and $15$. This feature is missing in each
of the liquid integral equation scheme results, which predict a rounder shape of the second peak.
Similar features in the second peak of the structure factor have been discussed as possible freezing precursors, and as signatures
of short-ranged order in the liquid phase \cite{Schenk2002,Homo2007} (see also the related Ref.~\cite{Gapinski2014}).
To the best of our knowledge, the second peak shape feature is not observed in any of the usual liquid integral equation schemes that are formulated on the level
of pair-correlation functions. A similar feature (in the radial distribution function, however, and for $d=3$) has been reported in Ref.~\cite{Brader2008}, where a 
computationally more sophisticated integral equation scheme was solved that includes non-trivial triplet correlations. The implementation of such a scheme
for the four-dimensional fluids under investigation is beyond the scope of the present work.

Note from Fig.~\ref{fig:Sq_T2_5_and_7_0}, that the agreement between the MD simulation and RY-scheme structure factors is very good for higher temperatures
($T = 2.5 \varepsilon / k_B$ in the left panel of Fig.~\ref{fig:Sq_T2_5_and_7_0}, and $T = 7.0 \varepsilon / k_B$ in the right panel). For $T = 7.0 \varepsilon / k_B$,
the flattened second peak feature has practically disappeared in the MD simulation results, and the agreement to the RY scheme is almost perfect.

\section{Conclusions}\label{sec:conclusions}
We have demonstrated that the RY integral equation scheme predicts pair correlations in
homogeneous four-dimensional fluids of particles with soft repulsive interactions in very good agreement with
numerically accurate, but computationally expensive Molecular Dynamics simulations.
This finding, which is in line with the known excellent performance of the RY scheme for three-dimensional
fluids, qualifies the RY scheme as a numerically highly efficient method for calculating the structure input
that is needed for theories of dynamics, phase behaviour and vitrification in higher dimensions \cite{Schmid2010}.
Despite its overall very good accuracy, the RY scheme (as well as the PY and HNC schemes) fails to predict
the correct shape of the second peak in the static structure factor when the particle repulsion becomes very strong.
We expect that inclusion of nontrivial triplet correlations into a (thermodynamically partially self-consistent) liquid integral
equation scheme \cite{Brader2008} for arbitrary spatial dimensions could result in an improved ability of the theory to
reproduce the static structure factor, particularly around its second peak.

\section*{Acknowledgements}
It is our pleasure to thank Ram\'{o}n Casta\~{n}eda-Priego, Mauricio D. Carbajal-Tinoco and Cristian Vasile Achim for helpful
comments and discussions.
M.H. acknowledges support by a fellowship within the Postdoc-Program of the German Academic Exchange Service (DAAD).
J.H. acknowledges funding via the DFG Research Unit FOR 1394 ``Nonlinear Response to Probe Vitrification''.
M.H. and H.L. acknowledge funding within the ERC Advanced Grant INTERCOCOS (project number 267499).

\end{document}